\documentclass[preprint2]{aastex}

\slugcomment{To appear in ApJ}

\shorttitle{C$_{60}$ in RCB stars}
\shortauthors{Garc\'{\i}a-Hern\'andez et al.}

\begin{document}


\title{Are C$_{60}$ molecules detectable in circumstellar shells of R
Coronae Borealis stars?}


\author{D. A. Garc\'\i a-Hern\'andez\altaffilmark{1,2}, N. Kameswara
Rao\altaffilmark{3} and D. L. Lambert\altaffilmark{4}}


\altaffiltext{1}{Instituto de Astrof\'{\i}sica de Canarias, C/ Via L\'actea
s/n, 38200 La Laguna, Spain; agarcia@iac.es}
\altaffiltext{2}{Departamento de Astrof\'{\i}sica, Universidad de La Laguna (ULL), E-38205 La Laguna, Spain}
\altaffiltext{3}{Indian Institute of Astrophysics, Bangalore 560034, India; nkrao@iiap.res.in}
\altaffiltext{4}{W. J. McDonald Observatory. The University of Texas at
Austin. 1 University Station, C1400. Austin, TX 78712$-$0259, USA; dll@astro.as.utexas.edu}


\begin{abstract}
The hydrogen-poor, helium-rich and carbon-rich character of the gas around R
Coronae Borealis (RCB) stars has been suggested to be a site for formation of
C$_{60}$ molecules. This suggestion is not supported by observations reported
here showing that infrared transitions of C$_{60}$ are not seen in a large
sample of RCB stars observed with the Infrared Spectrograph on the Spitzer Space
Telescope. The infrared C$_{60}$ transitions are seen, however, in emission and
blended with PAH-features in spectra of DY Cen and possibly also of V854 Cen,
the two least hydrogen-deficient (hydrogen deficiency of only $\sim$10$-$100)
RCB stars. The speculation is offered that  C$_{60}$ (and the PAHs) in the
moderately H-deficient circumstellar envelopes may be formed by the
decomposition of hydrogenated amorphous carbon but fullerene formation is
inefficient in the highly H-deficient environments of most RCBs. 
\end{abstract}


\keywords{circumstellar matter --- astrochemistry --- stars: chemically
peculiar --- stars: white dwarfs --- infrared: stars}



\section{Introduction}

R Coronae Borealis stars (RCBs, hereafter) stars are a rare class of supergiants
whose atmospheres are extremely hydrogen-deficient - the H-deficiency ranges
from about a factor of 10$-$100 to at least 10$^{8}$  - with helium the most
abundant element and often the stars are carbon-rich (e.g., Lambert \& Rao
1994).  RCBs at unpredictable intervals form thick carbon dust clouds which if
they form above the Earth-facing stellar surface  can obscure the star causing a
decrease up to eight magnitudes in the visual band and a decline can last from a
few weeks to many months. 

The RCB's hydrogen deficiency together with the helium-  and carbon-rich
character of the gas may facilitate the formation of molecular and dust species
not seen in  circumstellar envelopes of normal stars.  In particular, these
envelopes  have been considered to be possible environments for the formation of
the buckminsterfullerene molecule C$_{60}$ (e.g., Goeres \& Sedlmayr 1992).  The
remarkable stability of C$_{60}$ against intense radiation, ionization, etc.
(e.g., Kroto 1987) reinforces the idea that fullerenes such as C$_{60}$ could be
present in the interstellar medium were they formed in and ejected from
circumstellar envelopes of one or more kinds of mass-losing star.  

Are RCB stars a source of C$_{60}$ molecules?  Early observations of three RCB
stars (R CrB, RY Sgr, and V854 Cen) at a resolution R=1000 around 8.6$\mu$m
searched for the 8.4$\mu$m C$_{60}$ feature and reported a negative result
(Clayton et al. 1995).  With the advent of the Spitzer Space Telescope, we
undertook a new search for C$_{60}$ around RCBs. Identification of the C$_{60}$
molecule can be done in the infrared domain, where there are  infrared
transitions centered at $\sim$7.0, 8.4, 17.4 and 18.8 $\mu$m according to
gas-phase laboratory spectroscopy (Frum et al. 1991; Nemes et al. 1994).  In
this paper, we report two results. The first is that fullerenes are not seen
around most RCBs. The second is that the C$_{60}$ IR transitions are present
along with transitions due to PAHs in the spectrum of DY Cen and possibly also
in the spectrum of V854 Cen, which are the two least H-deficient RCBs known.
Spitzer observations have very recently provided evidence for C$_{60}$ and
C$_{70}$  from planetary nebulae (PNe; Cami et al. 2010;
Garc\'{\i}a-Hern\'{a}ndez et al. 2010) and reflection nebulae (Sellgren et al.
2010). None of these environments is H-deficient. Although Cami et al. declare
their observations refer to a H-poor region of the PN Tc 1,  the
literature does not support their claim (see below). These detections of
fullerenes with our detection of C$_{60}$ from RCBs with a modicum of H suggests
that formation of fullerenes may require some H in addition to C.   

\section{Spitzer observations and infrared spectra}

We have recently conducted an infrared spectral survey with the Infrared
Spectrograph (IRS) on the Spitzer Space Telescope for a complete sample
($\sim$30) of RCB stars spanning a full range of hydrogen content, temperature
and composition. The spectral energy distributions ($\sim$0.4$-$40 $\mu$m) were
constructed for all RCB stars in this sample (see Garc\'{\i}a-Hern\'andez et al.
2011 for more details). Since here we are interested in the emission and
absorption features present in the 6-25 $\mu$m window, we interpolate between
several points in the dust continuum and subtract this baseline to provide the
residual spectrum (Figure 1), where the dust and gas features may be easily
identified. 

Residual spectra of those RCBs with a H-deficiency in excess of a factor of
about 10$^3$ (Garc\'{\i}a-Hern\'andez et al. 2011) show only a broad 
$\sim$6-10 $\mu$m emission feature which is attributable to C-C stretching modes
of amorphous carbon grains (Colangeli et al. 1995). The profile of this feature
may vary slightly from RCB-to-RCB but this variation may not entirely be
intrinsic to the circumstellar envelopes. Extension immediately longward of
10$\mu$m is sensitive to the adopted interstellar reddening which determines the
correction for the 9.7$\mu$m silicate absorption feature. Figure 2 shows the
average residual spectrum for the sample (9) of the least reddened
(E$_{B-V}$$\sim$0.05$-$0.40) RCB stars (UW Cen, RT Nor, RS Tel, V CrA, V1157
Sgr, V1783 Sgr, S Aps, U Aqr, and Z Umi; Garc\'{\i}a-Hern\'andez et al. 2011). 

The RCB stars DY Cen and V854 Cen display apparently similar but very different
residual spectra from the majority of the RCBs (Figure 2).  DY Cen and V854 Cen
are the two stars in our sample that are the least H-deficient with
H-deficiencies of factors of $\sim$10 for DY Cen (T$_{eff}$=19500K; Jeffery \&
Heber 1993) and $\sim$ 100-1000 for V854 Cen (T$_{eff}$=6750K; Asplund et al.
1998).  DY Cen and V854 Cen display strong features at $\sim$6.3, 7.7, 8.6, and
11.3 $\mu$m together with weaker features at $\sim$11.9 and 12.7 $\mu$m, all of
which may be identified with polycyclic aromatic hydrocarbons (PAHs) (e.g.,
Allamandola et al. 1989; Bauschlicher et al. 2008, see Table 1). The PAH
features (with the exception of the strong 7.7 $\mu$m band) are quite narrow
($\sim$0.2$-$0.4 $\mu$m) showing that they probably arise from free gas-phase
PAHs, not PAHs in particles or clusters (Peeters et al. 2004). In addition, the
wavelength positions of the strong 7.7 $\mu$m band are 8.0 and 7.8 $\mu$m for
V854 Cen and DY Cen, respectively. The mean position of the 7.7 $\mu$m band is
correlated apparently with  effective temperature of the exciting star (see Fig.
5 in Tielens 2008): the 7.8 to 8.0 $\mu$m shift  between the two stars
is consistent with the difference in their stellar temperatures.

The intriguing result from Fig. 2 is the presence in the DY Cen residual spectrum
of three   features at $\sim$7.0, 17.4 and 18.8 $\mu$m  not readily attributable
to PAHs according to wavelength and/or intensity considerations.  These features,
which are real as they are detected in available low- and high-resolution Spitzer
spectra at all slit positions,  are attributable to C$_{60}$, as we show below. To
establish their carrier as C$_{60}$, it is necessary to discuss the anticipated
spectra of the fullerene and their possible blending PAHs.  Table 1 lists features
seen in DY Cen and V854 Cen as well as their identification.

\section{Identifications: C$_{60}$ and/or PAH?}

Laboratory gas-phase spectroscopy of neutral C$_{60}$ molecules was reported by
Frum et al. (1991) and Nemes et al. (1994).  Infrared C$_{60}$ features are
expected at $\sim$7.0, 8.4, 17.4 and 18.8 $\mu$m at a temperature of 0 K while
these wavelengths are predicted to shift longward a maximum of 0.2 $\mu$m at a
temperature of 1083 K (Nemes et al. 1994). There are apparently no reliable
estimates of the relative strengths of the four bands in the gas phase. Cami et
al. (2010) estimated Einstein A-values from published band absorption strengths
for C$_{60}$ in rare gas matrices: these values according to our calculations
are  A(s$^{-1}$) $\simeq$ (1.9, 1.1, 4.2, 5.2) for the bands (18.8, 17.4, 8.4,
7.0). Intensities of circumstellar features will depend also on the level
populations and radiative transfer effects. 

There is a plethora of astronomical, laboratory, and theoretical data on the
PAHs. Our primary argument applicable to DY Cen is that three of the four C$_{60}$
features are not blended beyond recognition with PAH features and that a
reasonable case may be made that the fourth feature at 8.4 $\mu$m is a blend of
the C$_{60}$ transition and the 8.6 $\mu$m PAH feature. 

In terms of upper state excitation energy, the C$_{60}$ bands are ordered by
decreasing wavelength and it is in this order that we discuss their presence in
DY Cen.  The two longest wavelength  C$_{60}$ features match emission features
in DY Cen at 17.40 $\mu$m and 18.98 $\mu$m in good agreement with laboratory
measurements at about 1000K and with their extrapolation to 0K; one expects the
temperature of the C$_{60}$ molecules to be within these limits. The
line widths (dominated by the instrumental width) are consistent with the
laboratory measurements, 0.31 and 0.36 $\mu$m (observed) versus 0.40 $\mu$m
(laboratory).  

In DY Cen's spectrum there are no other features between 15 and 30 $\mu$m  but
V854 Cen shows the 17.4 $\mu$m feature not only stronger than its 18.8 $\mu$m
counterpart but shifted to shorter wavelengths (17.33 $\mu$m) and
accompanied by emission extending to about 15 $\mu$m. These additional features
are presumed to be PAH features known to fall in the 13 to 19 $\mu$m interval
(Boersma et al. 2010). Except for a 16.4 $\mu$m feature, the intensities of
these PAH contributions are uncorrelated with the stronger PAH features in the 5
to 13 $\mu$m interval. Additionally, the relative intensities of features in the
13 to 19 $\mu$m region may vary from object to object. Often, the 16.4 $\mu$m
feature is the strongest in this window with an intensity correlated with that
of the 11.2 $\mu$m PAH. With a slight extrapolation of Boersma et al's Fig. 6
showing a correlation between the intensity ratio of 6.2/11.2$\mu$m and
16.4/11.2$\mu$m PAH features, the 16.4$\mu$m feature is predicted to have an
intensity 2\% that of DY Cen's 11.2$\mu$m feature, an expectation consistent
with the feature's absence (Fig. 2) and, since other PAH features in this
interval are expected to be weaker, it is not surprising that the C$_{60}$
longest wavelength transitions appear uncontaminated by PAH blends. For V854
Cen, the 6.2/11.2$\mu$m intensity ratio (Figure 2) is higher than for DY Cen and
predicts an intensity of about 5\% for the 16.4$\mu$m PAH, a value approximately
consistent with the presence of the 16.4$\mu$m and other weak PAH features in
Fig. 2. With a correction for weak 15.8, 16.4, and 17.0 $\mu$m PAH
contaminants,  a C$_{60}$ feature at about 17.4 $\mu$m is obtained that is
consistent with that of the 18.8 $\mu$m feature. 

Interestingly, DY Cen shows a unique spectrum across the 15 to 20 $\mu$m
interval among spectra exhibiting PAH features. Other features at 15.8, and 16.4
$\mu$m are not seen, as they are, for example, in reflection nebulae such as NGC
7023 (Sellgren et al. 2007, 2010; Boersma et al. 2010). Several PAHs from
the Ames spectral database show features near 17.4 and 19 $\mu$m but they are
always accompanied by other stronger features (e.g., at 16.4 $\mu$m:
Bauschlicher et al. 2008) but these features are absent from DY Cen's spectrum.
Some may be present in V854 Cen's spectrum. Sellgren et al. (2010) show that
the 17.4 $\mu$m feature has two components: one that correlates with the 18.9
$\mu$m C$_{60}$ feature and another one correlating with the 16.4 $\mu$m PAH
feature. The absence of the 16.4 $\mu$m PAH feature from the DY Cen spectrum
indicates that the 17.4 $\mu$m feature is dominated by C$_{60}$ emission.
However, in the case of V854 Cen since both 16.4 and 18.9 $\mu$m features are
seen, the 17.4 $\mu$m feature is due to a combination of C$_{60}$ and PAH
emission.

The 7.0 and 8.4 $\mu$m C$_{60}$ transitions are in the interval spanned by
common PAH features at 6.2, 7.7, and 8.5 $\mu$m. Of particular
concern is the blending of the 8.4$\mu$m C$_{60}$ and 8.5$\mu$m PAH features
where the PAH's contribution can be assessed only by comparison of relative
strengths of this and adjacent PAH features, an uncertain exercise owing to
considerable source-to-source variation in relative strengths.

The 7.0$\mu$m C$_{60}$ line is partially resolved in the DY Cen spectrum with a
measured wavelength of 7.0 $\mu$m in good agreement with the laboratory
spectroscopic value of 7.11 $\mu$m for a temperature of about 1000K. An estimate
of the feature's width requires a correction for the overlapping wing of the
strong 7.7$\mu$m PAH feature: a rough upper limit for the FWHM of the C$_{60}$
line is $<$0.16 $\mu$m, a value consistent with the laboratory measurement of
0.06 $\mu$m. In addition, we estimate an intensity ratio 7.0/18.9 of
$\sim$0.7, which - according to Sellgren et al. (2010) - corresponds to
excitation of C$_{60}$ molecules by photons with energies slightly less than 10
eV, in agreement with the effective temperature of DY Cen. The 8.4$\mu$m
C$_{60}$ line is blended with the 8.5$\mu$m PAH. There seems to be no way in
which to make a firm estimate of the PAH's contribution to the feature seen in
DY Cen's spectrum. Cerrigone et al. (2009) provide intensities for the principal
PAH features for both C-rich and O-rich post-AGB stars. Relative to the
11.2$\mu$m PAH, the mean intensity of the 8.5$\mu$m PAH is 0.40 for the four
C-rich stars (range 0.19 to 0.91), and 0.38 for the entire sample of 13 stars
(range 0.04 to 0.91). At the relative intensity of 0.4, the 8.5$\mu$m PAH is
expected to have an intensity close to the observed value. However, with respect
to the 6.2$\mu$m PAH, the mean intensity of the 8.5$\mu$m feature is 0.33 for
the four C-rich stars (range 0.23 to 0.5) and 0.19 for the entire sample (range
0.05 to 0.5). With these mean intensities, the predicted intensity of the
8.5$\mu$m PAH is less than the observed value in the DY Cen spectrum. Different
average relative intensities among PAH features  very likely reflect the
different origins of the features (Table 1). Sellgren et al.'s (2010)
calculations indicate that the 7.0 and 8.4 $\mu$m C$_{60}$ features should be
roughly similar in intensity. The predicted intensity of the 8.5$\mu$m PAH from
the 11.2$\mu$m PAH would satisfy this expectation but the prediction from the
6.2$\mu$m PAH would not. However, the considerable star-to-star variation in
relative intensities of these PAHs rules out making a reliable separation of
C$_{60}$ and PAH contributions. 

Inspection of V854 Cen's spectrum in Fig.2 reveals some differences with  DY
Cen's spectrum in the 6 to 10 $\mu$m interval. In particular, the strong
7.7$\mu$m PAH shows obvious blending to longer wavelengths - this feature
is broader in V854 Cen than in DY Cen. This may result from intrusion by an
additional PAH but another possibility is that V854 Cen includes a contribution
from the broad feature seen in the more H-deficient RCBs. In Fig.3, we show the
V854 Cen spectrum and the RCB feature scaled to provide an approximate possible
fit. The profile of the RCB feature seems to include some contributions at 7 and
8.5$\mu$m. In contrast to V854 Cen, the DY Cen spectrum does not appear to be
contaminated by the RCB feature and additional contributions at 7.0 and 8.5
$\mu$m are required which are most probably attributable to C$_{60}$. 

\section{Discussion}

The original laboratory studies on the formation of fullerenes showed that
fullerenes are clearly favored in environments which are H-deficient (Kroto et
al. 1985; Kratschmer et al. 1990) and that H-poor conditions are a prerequisite
for efficient fullerene formation (de Vries et al 1993). Thus, in the
circumstellar envelopes of cool evolved stars having a normal H abundance (e.g.,
C-rich Asymptotic Giant Branch stars) and in dense interstellar clouds,
acetylene (C$_{2}$H$_{2}$) and its radical derivatives are believed to be the
precursors of complex C-based molecules such as PAHs, and fullerenes are
probably not formed (e.g., Cherchneff \& Cau 1999). However, in hydrogen-poor
but C-rich environments, fullerene molecules may be formed from the coalescence
of large monocyclic rings in the gas phase and PAHs are likely not formed as
possible intermediates (e.g., Cherchneff et al. 2000). Furthermore, more recent
laboratory studies show that at low temperatures ($<$ 1700 K) soot formation
proceeds through or involves the formation of PAH intermediaries while
fullerenes are involved at temperatures in excess of 3500 K (J\"{a}ger et al.
2009). This shows that the high temperatures rather than the H-poor conditions
may also be a defining factor for efficient fullerene formation. Given the ease
with which fullerenes are formed in a carbon and helium rich atmosphere in
laboratory experiments, it is puzzling that fullerenes do not seem to form in
great abundance in the carbon and helium rich environments of the very H-poor
RCB stars, but, in fact, fullerenes are detected only around stars containing
some hydrogen (this study and Garc\'{\i}a-Hern\'{a}ndez et al. 2010).
Evidently, fullerenes formation is inefficient in the highly H-deficient
RCB stars; fullerene destruction is expected to occur at a slow rate.

Our detection of C$_{60}$ around DY Cen and possibly also around V854 Cen occurs
in conjunction with the presence of PAHs. As we have mentioned above, high
temperature condensation (even in H-rich environments) will lead to efficient
formation of fullerenes (J\"{a}ger et al. 2009) and this may be relevant for
these RCB stars since the gas in which molecules and dust form will be much
hotter than in red giant star environments. However, laboratory of high
temperature condensates have shown that no PAHs are formed as intermediates
(J\"{a}ger et al. 2009). An alternative explanation for the simultaneous
presence of PAH and C$_{60}$ molecules is that they may be formed by the
decomposition of hydrogenated amorphous carbon (HAC)  (Scott et al. 1997a). 
Laser vaporization of HAC films produces a wide range of large aromatic carbon 
molecules including  PAHs and fullerenes such as C$_{60}$ (Scott et al.
1997a).  Indeed, the C$_{60}$ molecules do not dominate the mass distribution
of molecules seen in laboratory experiments (Scott et al. 1997a), an observation
qualitatively consistent with the infrared spectra of  DY Cen and V854 Cen.
The UV radiation field around these two RCB stars is unlikely intense enough to
cause HAC destruction. However, high velocity strong winds are typical in RCB
stars and the collisional environment (i.e., grain-grain collisions) of these
stars may lead to HAC vaporization. 

Interestingly, the Infrared Space Observatory's 1996 spectrum of V854 Cen
showed a correspondence with the laboratory emission spectrum of HAC at
773 K (see Lambert et al. 2001). These HAC features are weaker, even
absent, from our Spitzer spectrum and the C$_{60}$ and PAH features present in
V854 Cen's Spitzer spectrum are weaker or absent in the ISO spectrum (see
Fig. 4). Although the absolute flux level is different for both spectra, the 
dust continuum emission seems to be unchanged and it can be well fitted by a
blackbody at a temperature of $\sim$1000 K. Figure 4 displays the ISO and
Spitzer spectra of V854 Cen after the subtraction of the dust continuum 
emission at $\sim$1000 K. This contrast between ISO and Spitzer spectra
necessarily prompts the speculation  that the principal ingredient in the
circumstellar envelope  evolved from HAC grains to molecules such as the PAHs
and C$_{60}$\footnote{Note that V854 Cen underwent minima between the time
of the ISO and Spitzer spectra and thus the HACs seen with ISO were not  - in
all probability - the HACs that led to the fullerenes and PAHs in the Spitzer
spectrum.}. If so, formation of  molecules such as PAHs and fullerenes in the
circumstellar envelopes of the more H-rich RCB stars is a time-dependent
phenomenon. A certain concentration of hydrogen is presumably needed to form HAC
grains, which may be then destroyed by shocks in the circumstellar
envelope. One product of destruction of HAC grains may be C$_{60}$ molecules
which being hardy may survive for longer periods of time than the HAC grains and
PAHs. In RCBs with a recurring series of dust-forming events with replenishment
of HACs and additional formation of C$_{60}$  is a possibility.  If these
speculations have merit, one may expect to find C$_{60}$ molecules unaccompanied
by HACs and PAHs in environments where grain formation is a non-recurring
event.  

Contrary to a conclusion drawn by Cami et al. (2010), our speculations may
account well for Cami et al.'s pioneering  detection of C$_{60}$ and C$_{70}$
molecules from the inner region of the PN Tc 1.  Noting that the  Spitzer
spectrum of Tc 1 shows no PAHs, Cami et al. drew the  conclusion that the
fullerenes were formed in very H-poor gas ejected a few thousand years by an AGB
star following an even earlier ejection of the star's H-rich envelope. This
conclusion overlooks two key observations. First, the nebula is not H-poor
(K\"{o}ppen et al. 1991; Milanova \& Kholtygin 2009) and, in particular, optical
and ultraviolet spectroscopy  of the inner regions confirm that the gas has a
normal mix of H and He (Williams et al. 2008;  R. Williams - private
communication); the gas is not H-poor. Second, the central star CoD -46$^\circ$
11816  is not H-poor and He-rich (Mendez 1991). In short, the fullerenes were in
all probability formed in H-rich and presumably C-rich gas. Dust formation
occurred in the circumstellar wind but now presumably the wind has lessened or
ceased.  The hot central star may have already destroyed the less hardy grains
and molecules (e.g., PAHs) from times when a cool AGB star fed the then stronger
wind.  In this picture, hydrogen is essential to form fullerenes but the
absence of PAHs is not proof that fullerene production occurs in a H-deficient
region.

\section{Concluding remarks}

In summary, contrary to general expectation, the formation of large fullerenes,
specifically C$_{60}$ molecules around RCB stars takes place efficiently only in
the presence of some hydrogen and HACs may be the precursors of
fullerenes. However, the absence of fullerene features in highly
H-deficient RCB stars is puzzling. Carbon chemistry in H-deficient environments
should include the formation of fullerenes, as it has been shown by the early
laboratory experiments (e.g., Kroto et al 1985; Kratschmer et al. 1990; de Vries
et al. 1993). More laboratory experiments at different temperatures and hydrogen
compositions are encouraged in order to learn about the formation of fullerenes.
In particular, laboratory experiments in H-poor atmospheres could explore higher
temperature formation routes of fullerenes. In addition, one would expect that
grain-grain collisions of pure carbon grains would lead to fullerenes, and in
this sense, laser vaporization experiments of amorphous carbon films could help
to solve this puzzle.



\acknowledgments

We thank the anonymous referee for useful comments that help to improve
this manuscript. We would like to thank Jack Baldwin and Rob Williams for their
quick clarification about Tc 1. This work is based on observations made with
the Spitzer Space Telescope, which is operated by the Jet Propulsion Laboratory,
California Institute of Technology, under NASA contract 1407.  D.A.G.H.
acknowledges support by the Spanish Ministry of Science and Innovation (MICINN)
under a JdC grant and under grant AYA-2007-64748. D.L.L. acknowledges support
for this work provided by NASA through an award for program GO 50212 issued by
JPL/Caltech. D.L.L. also wishes to thank the Robert A. Welch Foundation of
Houston, Texas for support through grant F-634. 



{\it Facilities:} \facility{Spitzer:IRS}.

\clearpage

\begin{deluxetable}{lccccc}
\tabletypesize{\scriptsize}
\tablecaption{Mid-infrared features  in DY Cen's spectrum.
\label{tbl-1}}
\tablewidth{0pt}
\tablehead{
\colhead{Feature} &   \colhead{DY Cen} & \colhead{V854 Cen}
&\colhead{Identification} & \colhead{Mode} & \colhead{Ref.$^{a}$}\\
}
\startdata
6.3 $\mu$m    &  yes  &  yes       & PAHs         & C-C stretching   & 1 \\
7.0 $\mu$m    &  yes  &  $\dots$   & C$_{60}$     & F$_{1u}$(4)      & 2 \\
7.7 $\mu$m    &  yes  &  yes       & PAHs         & C-C stretching   & 1 \\
8.6 $\mu$m    &  yes  &  $\dots$   & C$_{60}$, PAHs   & F$_{1u}$(3), C-H bending in-plane      & 2,1 \\
11.3 $\mu$m   &  yes  &  yes       & PAHs         & C-H bending out-of-plane & 1 \\
11.9 $\mu$m   &  yes  &  yes       & PAHs         & C-H bending duo  & 1 \\
12.7 $\mu$m   &  yes  &  yes       & PAHs         & C-H bending trio & 1 \\
15.8 $\mu$m   &  no   &  yes       & large PAHs?  & C-C-C            & 3 \\
16.4 $\mu$m   &  no   &  yes       & large PAHs?  & C-C-C            & 3 \\
17.0 $\mu$m   &  no   &  yes       & large PAHs?  & C-C-C            & 3 \\
17.4 $\mu$m   &  yes  &  yes$^{b}$ & C$_{60}$     & F$_{1u}$(2)      & 2 \\
18.8 $\mu$m   &  yes  &  yes       & C$_{60}$     & F$_{1u}$(1)      & 2 \\
\enddata
\tablenotetext{a}{References for the identification of the mid-IR features.} 
\tablenotetext{b}{Note that the 17.4 $\mu$m feature seen in V854 Cen is due to a combination of C$_{60}$ and PAH
emission (see text for more details).}.
\tablerefs{(1) Allamandola et al. (1989); (2) Frum et al. (1991); (3) Boersma et
al. 2010.}
\end{deluxetable}

\clearpage
 \begin{figure}
\includegraphics[angle=0,scale=.60]{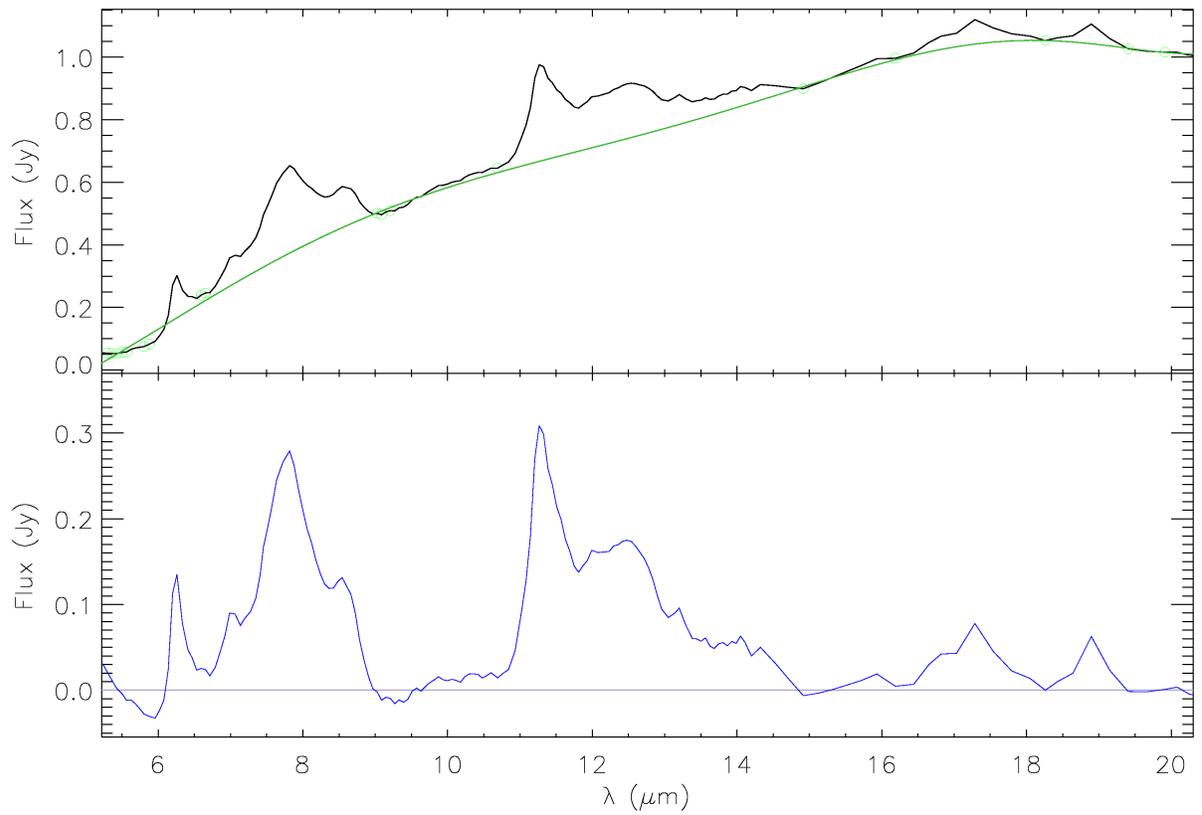}
\caption{Observed Spitzer/IRS spectrum of DY Cen (in black) together with a
polynomial fit (in green) to continuum points free grom any gas and dust
feature. The corresponding residual spectrum (in blue) is shown in the bottom
panel.\label{fig1}}
\end{figure}

\clearpage

\begin{figure}
\includegraphics[angle=0,scale=.60]{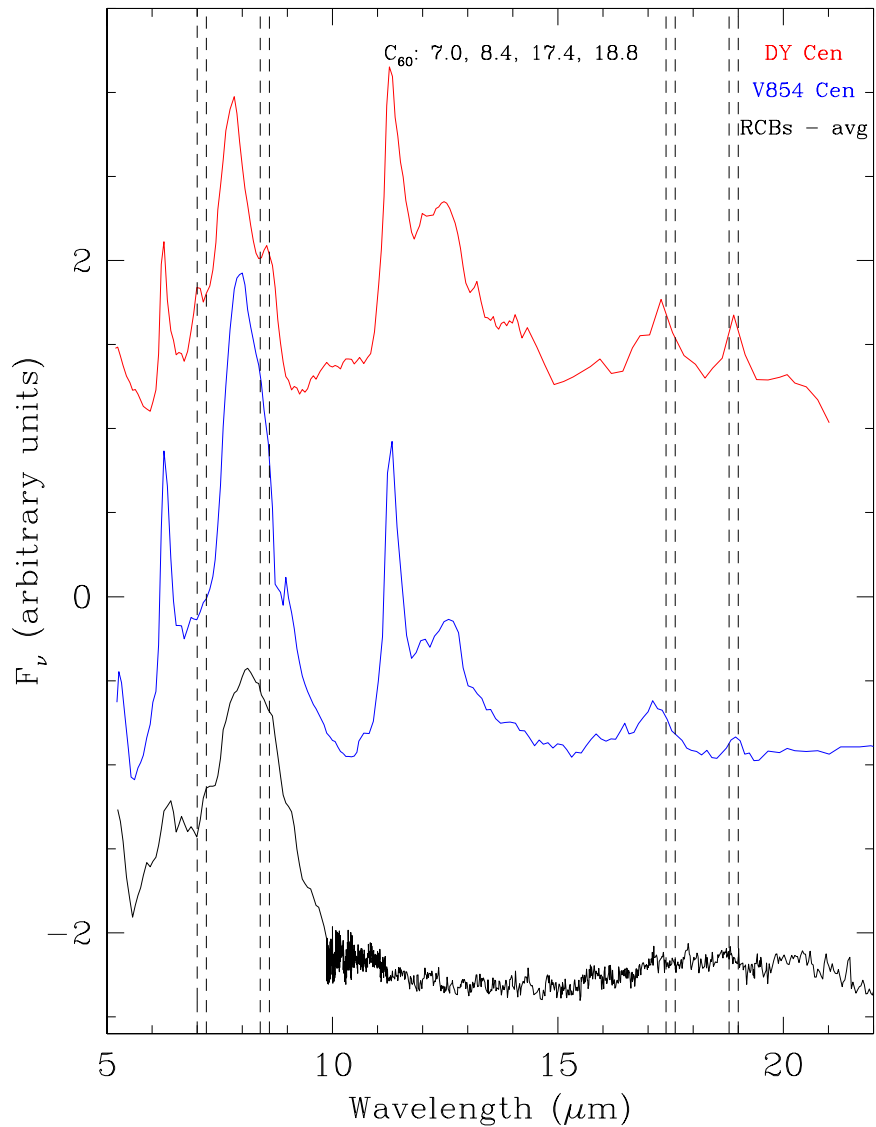}
\caption{Spitzer/IRS residual spectra in the wavelength range $\sim$5$-$20
$\mu$m of the RCB stars DY Cen (in red) and V854 Cen (in blue). The average
residual spectrum of nine extremely H-deficient RCBs with little reddening (in
black) is also shown. The expected temperature-dependent positions of the
neutral C$_{60}$ features are marked with black dashed vertical lines. 
\label{fig2}}
\end{figure}

\clearpage

\begin{figure}
\includegraphics[angle=0,scale=.60]{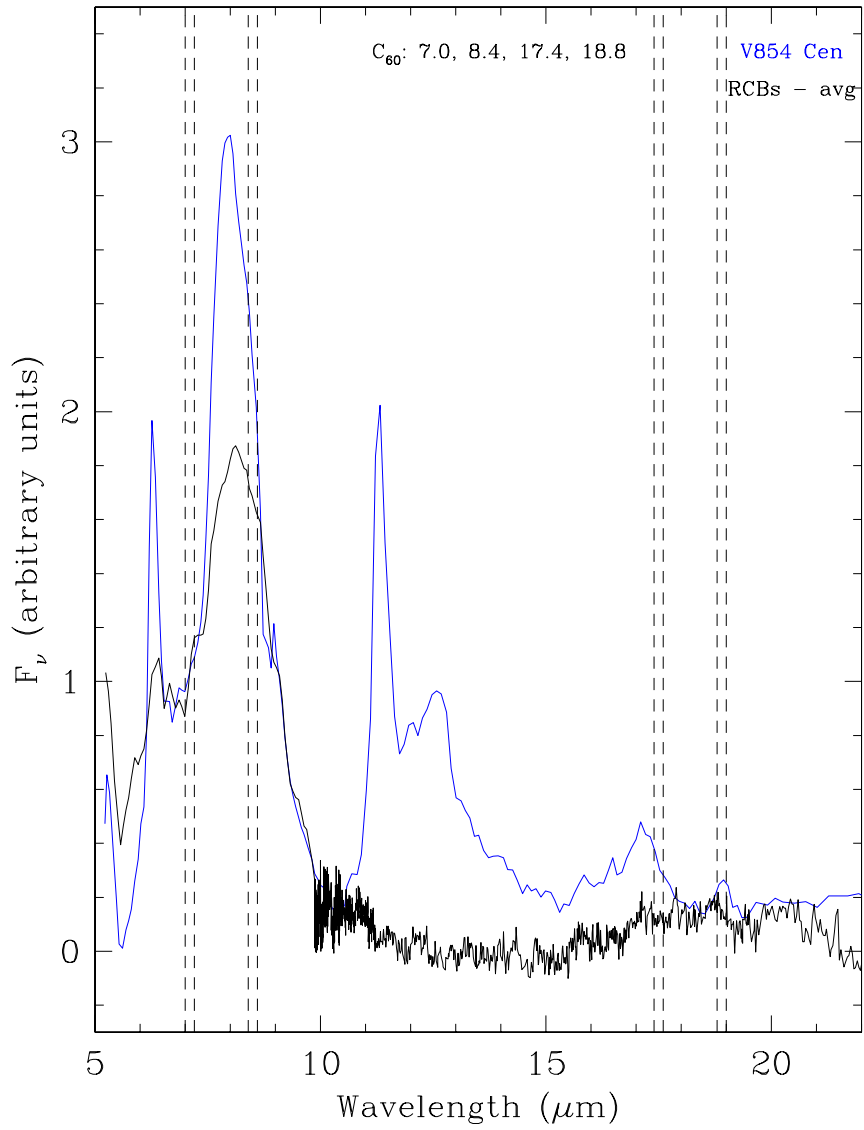}
\caption{V854 Cen residual spectrum compared with the average residual
RCB spectrum. \label{fig3}}
\end{figure}

\clearpage

\begin{figure}
\includegraphics[angle=0,scale=.60]{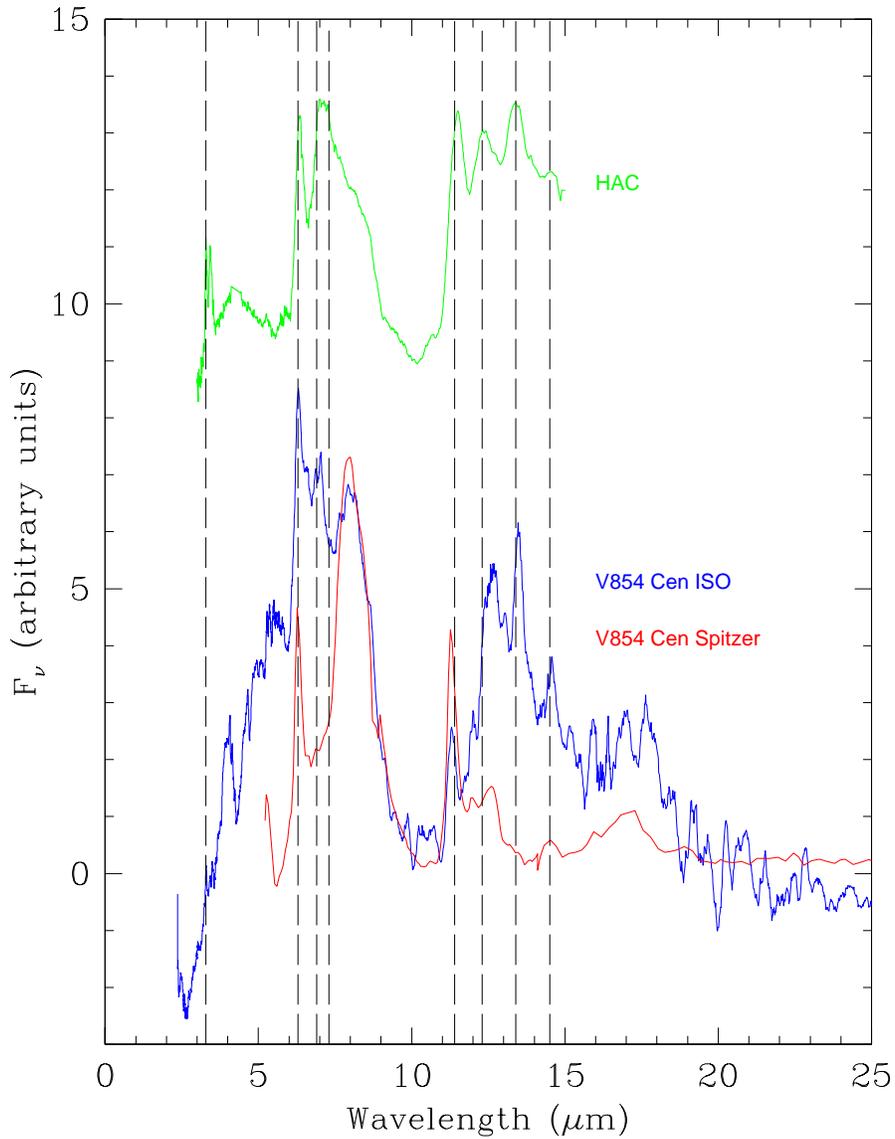}
\caption{Residual ISO 1996 September 9 (in blue) and Spitzer/IRS 2007 September
7 (in red) spectra in the wavelength range $\sim$2-25 $\mu$m for the RCB star
V854 Cen. A blackbody of $\sim$1000 K was subtracted from both spectra. The ISO
spectrum (R$\sim$1000) has been smoothed with a 13 box car in order to be
compared with the Spitzer spectrum. The laboratory emission spectrum of HAC at
773 K (in green; Scott et al. 1997b) is shown for comparison. The main
laboratory HAC emission features (Scott et al. 1997b) are marked with black
dashed vertical lines. \label{fig4}}
\end{figure}

\end{document}